\documentclass[aps,pre,superscriptaddress,twocolumn,letterpaper]{revtex4}
\usepackage{amsfonts,amssymb,amsmath,bm,bbm,mathrsfs}
\usepackage[usenames]{color}
\usepackage{color,soul}
\usepackage[utf8]{inputenc}
\usepackage[T1]{fontenc}
\usepackage{textcomp}
\usepackage{upgreek}
\usepackage{graphicx}
\usepackage[tight,sf]{subfigure}

\usepackage{hyperref}
\newcommand{\arcsinh}{\mathrm{arcsinh} \,}
\newcommand{\rmd}{\mathrm{d}}

\newcommand{\frk}{\mathfrak{K}}
\newcommand{\frm}{\mathfrak{m}}

\newcommand{\calH}{\mathcal{H}}
\newcommand{\bfx}{\mathbf{x}}
\newcommand{\bfy}{\mathbf{y}}
\newcommand{\bfe}{\mathbf{e}}
\newcommand{\rme}{\mathrm{e}}
\newcommand{\bfn}{\mathbf{n}}
\newcommand{\rmn}{\mathrm{n}}
\newcommand{\bfX}{\mathbf{X}}
\newcommand{\bft}{\mathbf{t}}
\newcommand{\rmt}{\mathrm{t}}
\newcommand{\bfl}{\mathbf{l}}
\newcommand{\rml}{\mathrm{l}}
\newcommand{\bff}{\mathbf{f}}
\newcommand{\bfm}{\mathbf{m}}
\newcommand{\hbfz}{\hat{\bf z}}
\newcommand{\bfF}{\mathbf{F}}
\newcommand{\bfM}{\mathbf{M}}
\newcommand{\bfH}{\mathbf{H}}

\begin{document}

\title{Flexible paramagnetic membranes in fast precessing fields}

\author{Pablo V\'{a}zquez-Montejo}
\email{pablo.vazquez@correo.uady.mx}
\affiliation{Department of Materials Science and Engineering, Northwestern University, 
2220 Campus Drive, Evanston, Illinois, 60208, USA}
\affiliation{Facultad de Matemáticas, Universidad Autónoma de Yucatán, 
Periférico Norte, Tablaje 13615, Mérida, Yucatán, 97110, MÉXICO}
\affiliation{Instituto de Matemáticas, Universidad Nacional Autónoma de México, 
Circuito exterior, Ciudad Universitaria, Ciudad de México, 04510, MÉXICO}
\author{M\'{o}nica Olvera de la Cruz}
\email[]{m-olvera@northwestern.edu}
\affiliation{Department of Materials Science and Engineering, Northwestern University, 
2220 Campus Drive, Evanston, Illinois, 60208, USA}
\affiliation{Department of Physics and Astronomy, Northwestern University,
2145 Sheridan Road F165, Evanston, Illinois, 60208, USA}

\begin{abstract}
Elastic membranes composed of paramagnetic beads offer the possibility of assembling versatile actuators operated autonomously by external magnetic fields. Here we develop a theoretical framework to study shapes of such paramagnetic membranes under the influence of a fast precessing magnetic field. Their conformations are determined by the competition of the elastic and magnetic energies, arising as a result of their bending and the induced dipolar interactions between nearest neighbors beads. In the harmonic approximation, the elastic energy is quadratic in the surface curvatures. To account for the magnetic energy we introduce a continuum limit energy, quadratic in the projections of the surface tangents onto the precession axis. We derive the Euler-Lagrange equation governing the equilibria of these membranes, as well as the corresponding stresses. We apply this framework to examine paramagnetic membranes with quasiplanar, cylindrical, and helicoidal geometries. In all cases we found that their shape, energy, and stresses can be modified by means of the parameters of the magnetic field, mainly by the angle of precession. 
\end{abstract}


\maketitle

\section{Introduction}

Magnetic fields can be used to direct magnetic particles in multiple environments including living tissues. Since most materials are not magnetic, magnetic particles can be directed in media where variables such as  temperature and chemical composition cannot be controlled.  For this reason they are extensively used in biotechnology and medicine \cite{Kozissnik2013, Duran2008}.  Moreover, magnetic fields can be altered much faster than colloidal diffusion timescales. These features give magnetic particles  possibilities to direct self-assembly \cite{Wang2013, Dempster2015} or to synthesize magnetic gels or magnetic elastomers, which can be used for actuation or transport \cite{Weeber2017}, as well as to be exploited in the design of programmable robots able to perform tasks at small scales \cite{Hu2018, Kim2018}.  Indeed,  paramagnetic filaments, synthesized by joining paramagnetic beads with semiflexible polymers, have found diverse applications such as micromechanical sensors or self-propelled swimmers \cite{Biswal2003, Goubault2003, Dreyfus2005, Roper2008, Cebers2015}. They are highly versatile. For instance, depending on their rigidity and the magnetic field strength, arrays of paramagnetic filaments may collapse into hairpins, loops, sheets, or pillars \cite{Wei2016}, and in a fast precessing magnetic field, depending on the precession angle they may adopt planar or helical conformations \cite{Dempster2017, Vazquez2017}.
\\
An obvious extension in the study of this kind of systems would be the consideration of the two-dimensional counterparts of paramagnetic filaments, that is, membranes consisting of two-dimensional arrays of paramagnetic beads linked by elastic polymers. However, such systems have not been synthesized yet and previous work on two-dimensional arrays of magnetic nanoparticles has been devoted to free magnetic colloids confined to a plane \cite{Muller2014}, or to arrays of magnetic filaments on a plane \cite{Domingos2017}. Here we study theoretically such paramagnetic membranes in a fast precessing magnetic field, whose energy is associated with their bending and the interaction between the induced dipoles on the beads. To address the latter contribution, we introduce an energy density describing the continuum limit of the dipolar interactions between nearest neighbors, analogous to the linear magnetic energy density for paramagnetic filaments \cite{Goubault2003, Cebers2003, Vazquez2017}. Although general frameworks for studying elastic membranes in a magnetic field have been developed using different approaches, for instance magnetoelastic theory \cite{Steigmann2004, Chatzigeorgiou2015}, here we determine their equilibrium configurations from a variational principle that exploits the geometric character of the membrane's energy. In this framework, the stresses are expressed in terms of the membrane geometry, and the Euler-Lagrange (EL) equation is given by the conservation of the  stresses along the normal of the membrane \cite{CapoGuv2002, Guven2004, Guven2018}. While the EL equation and the stress tensor for the purely elastic case are well-known \cite{OuYang1989, CapoGuv2002, Guven2004, Deserno2015, Guven2018}, here we present their magnetic counterparts. We apply this framework to analyze paramagnetic membranes with almost-planar, cylindrical, and helicoidal shapes, along with the forces required to hold them.

\section{Membrane energy and stresses}

\begin{figure}[htbp]
\includegraphics[width=0.48\columnwidth]{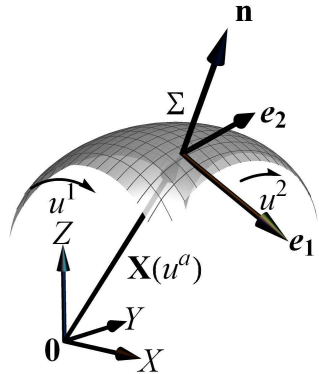}
\caption{The surface $\Sigma$ is parametrized by local coordinates $u^1$ and $u^2$; the tangent and unit normal vectors are $\bfe_a$ and $\bfn$.}	\label{fig:1}
\end{figure}

The paramagnetic membrane is represented by the surface passing through the centers of the beads, parametrized by two local coordinates $u^a$, $a=1,2$ in three-dimensional Euclidean space as $\Sigma: u^a \rightarrow \bfX(u^a) \in \mathbb{E}^3$. The tangent vectors adapted to this parametrization and the unit normal vector are $\bfe_a = \partial_a \bfX$ ($\partial_a = \partial/\partial u^a$) and $\bfn = \bfe_1 \times \bfe_2/\lVert \bfe_1 \times \bfe_2 \rVert$, as illustrated in Fig. \ref{fig:1}.
The components of the metric and extrinsic curvature tensors are defined by $g_{ab} = \bfe_a \cdot \bfe_b$ and $K_{ab} =  \bfe_a \cdot \partial_b \bfn$. The determinant of the metric is denoted by $g$. Indices are raised with the inverse of the metric $g^{ab}$. The shape operator $K^{a}_{\phantom{a}b} = g^{ac} K_{cb}$ is a linear map acting on surface tangent vectors and whose eigenvectors and eigenvalues correspond to directions and values of minimum and maximum curvature, known as principal directions and principal curvatures $C_1$ and $C_2$, respectively. The two invariants of $K^{a}_{\phantom{a}b}$ are its trace $K=C_1+C_2$ and determinant $K_G=C_1 C_2$ known as (twice) the mean and Gaussian curvatures \cite{doCarmoBook, FrankelBook}. 
\\
The energy ascribed to the membrane has two contributions that depend on its geometry. The first one, due to its bending, is given by the Canham-Helfrich energy density, quadratic in the curvatures \cite{Canham1970, Helfrich1973, Evans1974},
\begin{equation} \label{def:HB}
\calH_B[\bfX] = \frac{\frk}{2} \, K^2 + \frk_G \, K_G\,,
\end{equation}
where $\frk$ and $\frk_G$ are the bending and Gaussian moduli (with units of energy). We consider membranes composed by a single layer of paramagnetic beads, so we do not include a spontaneous curvature in the first term.
\\
The second energy contribution, is given in the quasistatic regime by the time-averaged dipolar interaction between nearest neighbors, induced by a magnetic field $\bfH$ precessing at an angle $\vartheta$ about an axis, which we choose as the $Z$ axis (further details are presented in Appendix \ref{App:derHM}), as shown in Fig. \ref{fig:2}.
\begin{figure}[htbp]
\includegraphics[width=0.44\columnwidth]{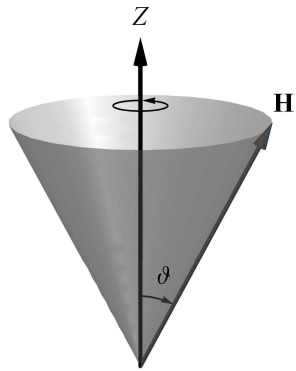}
\caption{The axis of precession of the magnetic field is chosen as the $Z$ axis and the angle of precession is $\vartheta$.}	\label{fig:2}
\end{figure}
This magnetic energy density is quadratic in the projections of the tangent vectors  onto the precession axis,
\begin{equation} \label{eq:HM}
\calH_M[\bfX]  = -\frac{\frm}{2} g^{ab} \rme_a^z \rme_b^z \,, \quad \rme_a^z = \bfe_a \cdot \hat{\bf z}\,,
\end{equation}
where the magnetic modulus $\frm$ (with units of energy per area) depends on the precession angle $\vartheta$,
\begin{equation} \label{def:magmodsurf}
\frm = \frac{\mu_0}{4 \pi \Delta l} \left(\frac{3 \mu}{\Delta l^2}\right)^2 \left(\cos^2 \vartheta - \frac{1}{3}\right) \,,
\end{equation}
with $\mu_0$ the vacuum permeability, $\mu$ the magnitude of the magnetic dipoles, and $\Delta l$ the separation between their centers. As we will see below, membrane conformations depend to a large extent on the precession angle through this magnetic modulus. $\frm(\vartheta)$ vanishes at the so-called ``magic'' angle $\vartheta_m = \arccos\, (1/\sqrt{3})$, and the leading order of the magnetic energy will be the quadrupolar term, which is of short range, so elasticity of the membranes will dominate \cite{Osterman2009}. In the regime $0 < \vartheta < \vartheta_m$ ($\vartheta_m < \vartheta <\pi/2$), we have $\frm(\vartheta)>0$ ($\frm(\vartheta)<0$), and from Eq. (\ref{eq:HM}) we see that to minimize $\calH_M$, membranes will tend to develop vertical (horizontal) regions aligned with (orthogonal to) the precession axis, so as to maximize (minimize) the projections $\rme_a^z$, analogous to the behavior of magnetic particle suspensions, which depending on the precession angle may arrange in chains or sheets along and orthogonally to the precession axis \cite{Martin2000}.
\\
We rescale all quantities by the bending modulus $\frk$ and denote the rescaled quantity by an overbar. In particular, the rescaled quantity $\bar{\frm} = \frm/\frk$ possesses dimensions of inverse area, so the square root of its inverse, $\ell = 1/\sqrt{|\bar{\frm}|}$, provides the characteristic length scale at which elastic and magnetic terms are comparable. On length scales smaller than $\ell$, elasticity dominates, and on length scales larger than $\ell$, magnetic effects dominate. The dimensionless parameter $\gamma = \bar{\frm} A = \mathrm{sign}(\frm) (A/\ell^2)$, which we call the magnetoelastic parameter in analogy with the quantity corresponding to filaments \cite{Cebers2003}, quantifies the ratio of magnetic to elastic energies ($H_B \sim \frk$ and $H_M \sim \frm A$, so $H_M/H_B \sim \gamma = \bar{\frm} A$), so for large (small) values of $\gamma$, magnetic (elastic) effects become dominant. Plausible experimental values of the magnetoelastic parameter are of the order $|\gamma| \approx 10^{-2} - 10^{4}$ (see Appendix \ref{App:gammavals}).
\\
The total energy of the membrane is given by $H[\bfX] = H_B + H_M$, where $H_B =\int \rmd A \calH_B$, and $H_M = \int \rmd A \calH_M$, are the total bending and magnetic energies, with $\rmd A = \sqrt{g} \, \rmd u^1 \wedge \rmd u^2$ the area element ($g= \det g_{ab}$). $H_B$ and $H_M$ are invariant under reparametrizations and translations; however, while $H_B$ is invariant under rotations, $H_M$ is only invariant under rotations about the precession axis.
\\
We assume that the paramagnetic membrane is inextensible, so we introduce a term fixing the total area $A_0$ and consider the effective energy $H_E = H + \sigma (A-A_0)$, where $\sigma$ is a Lagrange multiplier that can be regarded as an intrinsic surface tension ($\sigma >0$ represents tension and $\sigma<0$ compression). We do not consider a term fixing the total volume, because the array of paramagnetic beads is likely to have interstices. The variation of $H_E$ under a deformation of the surface $\delta \bfX$ is given by 
\cite{CapoGuv2002, Guven2004, Deserno2015, Guven2018}
\begin{equation} \label{eq:1stvarH}
\delta H_E =  \int \rmd A \nabla_a \bff^a \cdot \delta \bfX + \int \rmd A \nabla_a \delta Q^a \,,
\end{equation}
where $\nabla_a$ is the covariant derivative compatible with $g_{ab}$. The first term describes the response of the energy to a deformation in the bulk in terms of the divergence of the surface stress tensor $\bff^a = f^{ab}  \bfe_b + f^a \bfn$, whose components are defined by
\begin{subequations} \label{eq:vecF}
\begin{eqnarray} 
\bar{f}^{ab} &=&  K \left(K^{ab}- \frac{K}{2}g^{ab} \right) + g^{ab} \left( \frac{\bar{\frm}}{2} \, g^{cd} \, \rme_c^z \, \rme_d^z - \bar{\sigma} \right); \qquad \label{eq:fab}\\
\bar{f}^a &=&  g^{ab}  \left( -\nabla_b K + \bar{\frm} \, \rme_b^z \, \rmn^z \right) \,, \quad \rmn^z = \bfn \cdot \hbfz \,.
\end{eqnarray}
\end{subequations}
Note that for minimal surfaces with $K=0$, the elastic contributions vanish, but not the magnetic ones. 
The total force on a region of the membrane is given by $\mathbf{F} = \int \rmd s \bff_\perp$, where $\rmd s$ is the arc length along the boundary and $\bff_\perp = g_{ab} l^a \bff^b$, with $l^a$ the components of the outward conormal (further details on the projected stress tensor $\bff_\perp$ are presented in Appendix \ref{App:stressprojs}, whereas the surface torque tensor and the total torque on a region of the membrane can be found in Appendix \ref{App:torquetensor}).
\\
By using the Stokes theorem in the second term of Eq. (\ref{eq:1stvarH}), where
\begin{equation} \label{def:Qa}
\delta \bar{Q}^a = -\bar{\bff}^a \cdot \delta \bfX + \left[K g^{ab} - \bar{\frk}_G \left(K^{ab} - K g^{ab}\right)\right] \bfe_b \cdot \delta \bfn\,,
\end{equation}
it can be recast as a line integral along the boundaries. Thus, it represents the change of the energy due to a deformation of the boundaries, which will be useful to determine the boundary conditions. $\frk_G$ only enters the variation of the energy through $\delta Q^a$ (it does not feature in $\bff^a$), which is a consequence of the Gauss-Bonnet theorem \cite{doCarmoBook, FrankelBook, Deserno2015, Guven2018}. $H_M$ does not contribute to $\delta Q^a$.
\\
From Eq. (\ref{eq:1stvarH}) we see that stationarity of the energy implies the conservation of the stress tensor $\nabla_a \bff^a = \mathbf{0}$, consequence of the translational invariance of $H$ \cite{CapoGuv2002, Guven2004, Guven2018}.
The tangential projections of this conservation law vanish identically on account of the reparamerization invariance of $H$ \cite{CapoGuv2002, Guven2004, Guven2018}, whereas the normal projection, $\mathcal{E} : = \nabla_a \mathbf{f}^a \cdot \mathbf{n}$, provides the EL equation 
\begin{eqnarray} \label{eq:ELmagmemb}
\bar{\mathcal{E}}  &=& \left(-\Delta+2 {\cal K}_G - \frac{K^2}{2} + \bar{\sigma}\right) K   \nonumber \\
&+& \bar{\frm} \left( \left(K^{ab} -  \frac{K}{2} g^{ab}\right) \rme_a^z \, \rme_b^z - K \rmn^z{}^2\right) = 0 \,,
\end{eqnarray}
where $\Delta = g^{ab} \nabla_a \nabla_b$ is the Laplace-Beltrami operator. The elastic terms consist of the Laplacian of the mean curvature and a cubic function of the curvatures \cite{CapoGuv2002, Guven2004, Deserno2015, Guven2018}, whereas the magnetic terms are linear in the curvatures, but quadratic in the projections of the adapted basis onto the precession axis. 
\\
It is noteworthy to mention that Eq. (\ref{eq:HM}) is the simplest covariant expression of $\calH_M$ (equivalent expressions quadratic in $\rme_a^z$ are presented in Appendix \ref{App:altexp}). Moreover, we also point out that it is also possible to recast $\calH_M$, as well as the magnetic parts of $f^{ab}$ and $\mathcal{E}$, in terms of $\rmn^z{}^2$, such that it becomes apparent that $\calH_M$ is reparametrization invariant (details are presented in Appendix \ref{App:altexp}).

\section{Equilibrium configurations}

The EL Eq. (\ref{eq:ELmagmemb}) is of fourth order in derivatives of the embedding functions $\bfX$, which makes very difficult to solve it in general. However, we can resort to known solutions of the purely elastic case and examine how they get modified when the magnetic contribution is present. It is well known that for appropriate boundary conditions and material parameters, planes, cylinders, spheres,  Clifford tori, and minimal surfaces minimize the bending energy \cite{Seifert1997, Deserno2015, Guven2018}. We now analyze perturbative solutions of EL Eq. (\ref{eq:ELmagmemb}) about some of these geometries (relevant expressions of the required geometric quantities are collected in Appendix \ref{App:geomquant}). Let us begin with the simplest type.

\subsection{Almost planar membranes}

\begin{figure}[htb]
\subfigure[$n=m=1$]{\includegraphics[width=0.475\columnwidth]{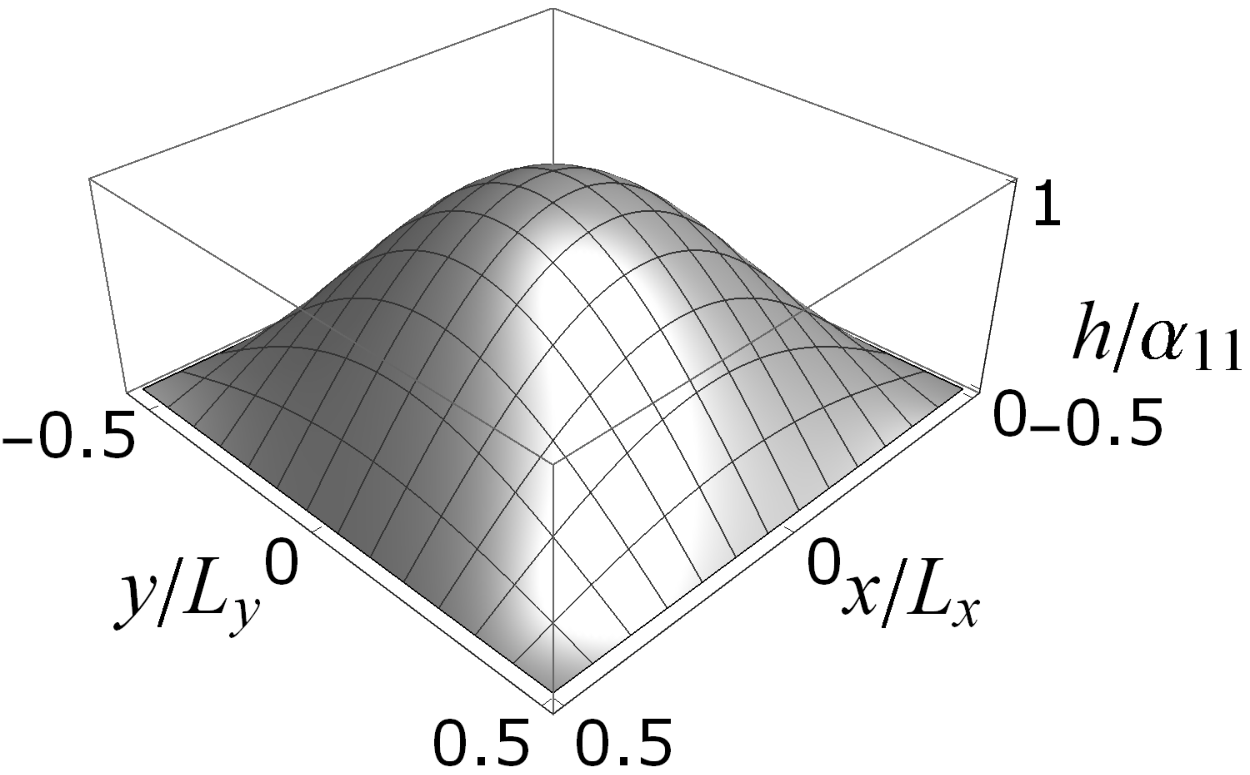}}
\subfigure[$n=1$, $m=2$]{\includegraphics[width=0.475\columnwidth]{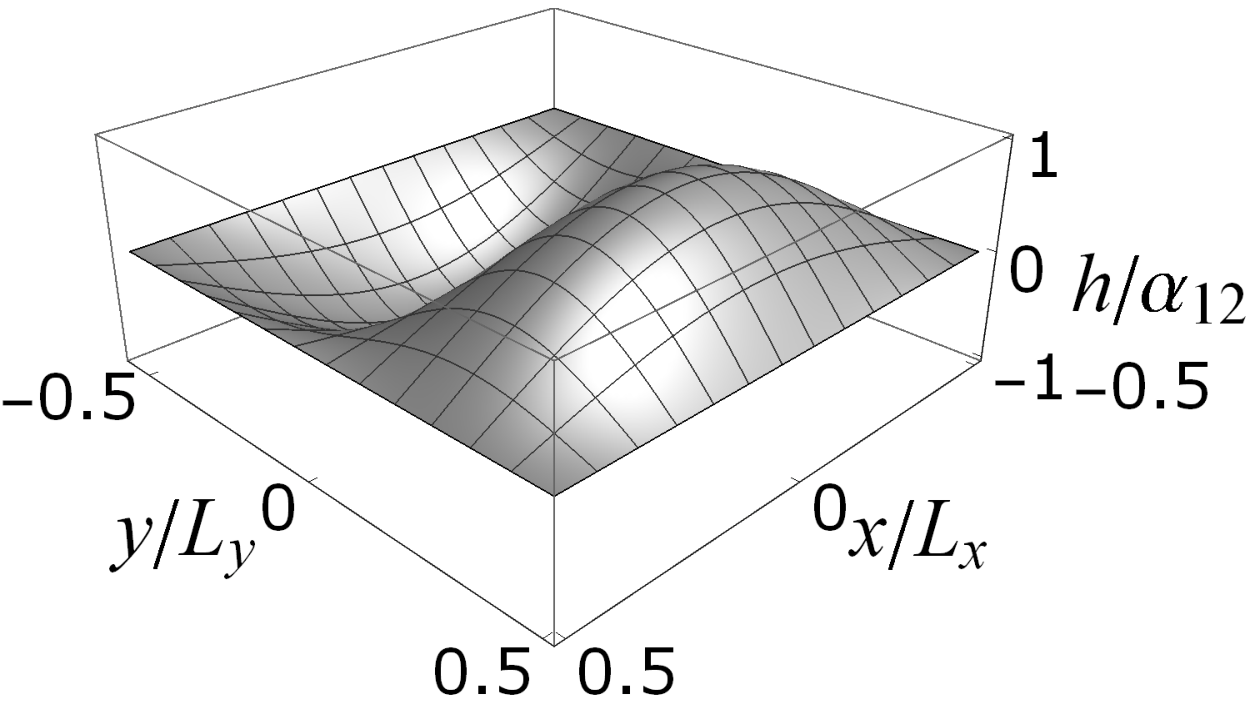}}\\
\subfigure[$n=m=2$]{\includegraphics[width=0.475\columnwidth]{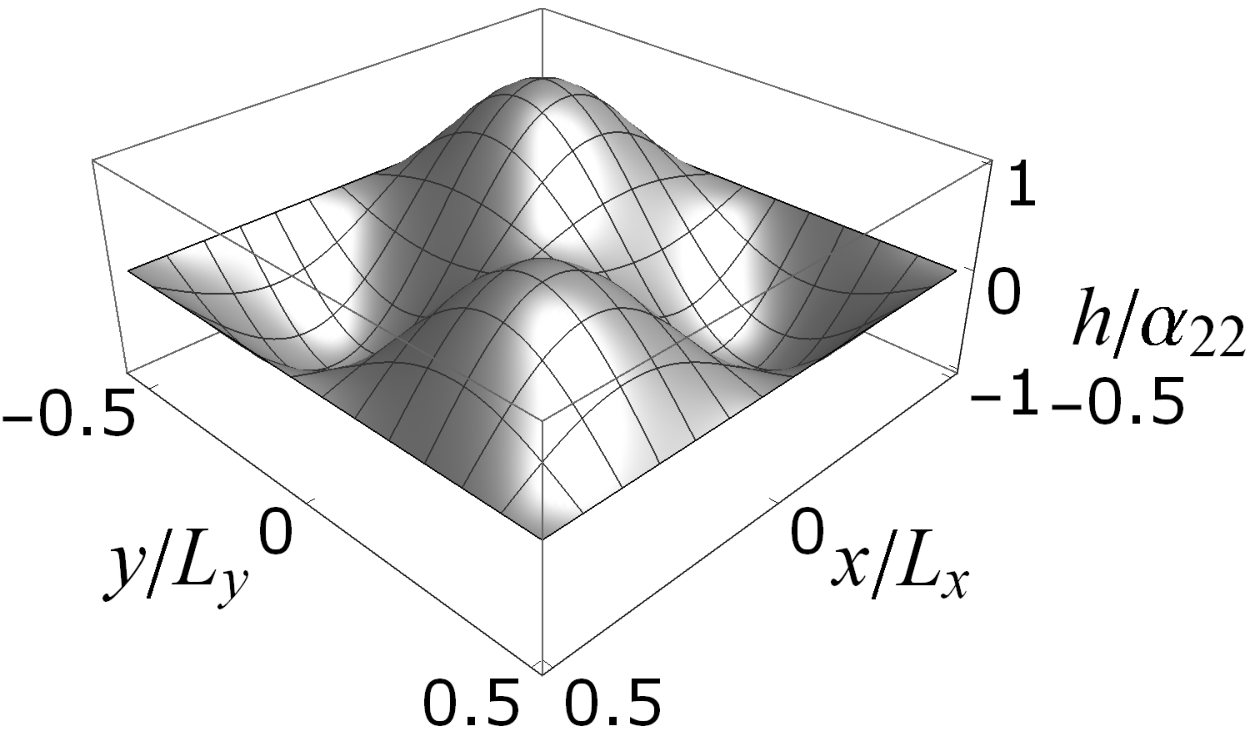}}
\subfigure[$n=2$, $m=3$]{\includegraphics[width=0.475\columnwidth]{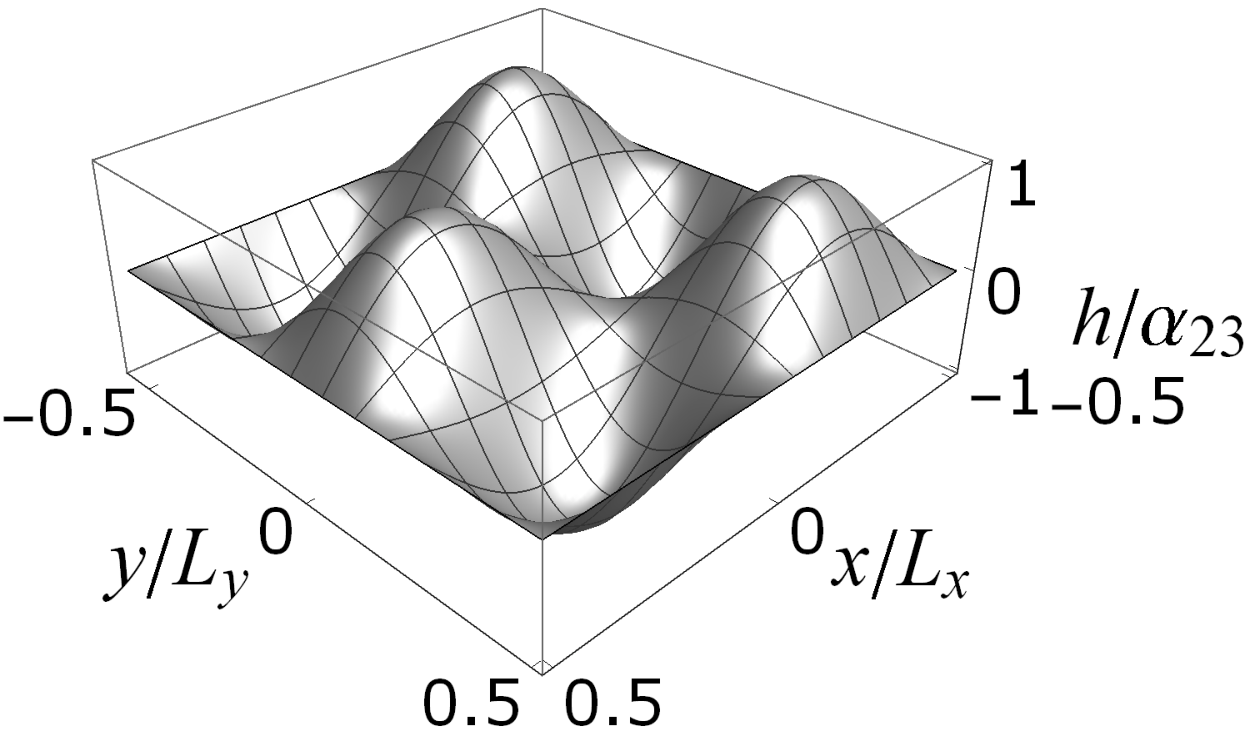}}
\caption{(a) Ground state and (b)-(d) first excited states resulting from the compression of a planar membrane.} \label{fig:3}
\end{figure}

Consider the deformation of a planar membrane orthogonal to the precession axis. It can be described in the Monge representation by a height function $z=h(x,y)$, parametrized by cartesian coordinates $x$ and $y$ on the base plane. For simplicity, we examine linearized solutions about the base plane, i.e., solutions in the small gradient approximation $|\nabla h| \ll 1$, where $\nabla h= \partial_x h \hat{\bfx} + \partial_y h\hat{\bfy}$. In this regime we have that up to second-order the projections of the adapted basis onto the precession axis are $\rme_a^z = \partial_a h$, and $\rmn^z \approx 1- (\nabla h)^2/2$. Furthermore, the mean and Gaussian curvature are given by $K \approx -\nabla^2 h$, with $\nabla^2 = \partial^2_x + \partial_y^2$ the Laplacian on the base plane, and $K_G \approx \partial^2_x h \partial^2_y h - (\partial^2_{xy} h)^2$ \cite{Fournier2007, Deserno2015, Guven2018}. Hence, at lowest order in derivatives of $h$, the EL Eq. (\ref{eq:ELmagmemb}) reduces to the Helmholtz equation $\bar{\mathcal{E}} \approx -(\nabla^2 +\bar{\Sigma})K=0$, with $\Sigma = - \sigma + \frm$ (recall that under compression $\sigma<0$). We see that the magnetic modulus renormalizes the intrinsic surface tension, and depending on the sign of $\frm$, it can be augmented or reduced. Trivial solutions of the EL Eq. (\ref{eq:ELmagmemb}) are given by minimal surfaces with $K =0$, which in this regime are described by  harmonic functions satisfying $\nabla^2 h =0$ \cite{Guven2018}. Another set of solutions is provided by the Helmholtz equation for $h$, $(\nabla^2 + \bar{\Sigma})h=0$, so in this regime the mean curvature is proportional to the height function, $K \approx \bar{\Sigma} h$. Let us consider a rectangular membrane with fixed edges of lengths $L_x$ and $L_y$ and of total area greater than its projected area, $A > A_p = L_x L_y$. From Eq. (\ref{def:Qa}) follows that stationarity of the energy requires the vanishing of the mean curvature at boundaries, so $K(0,y)=K(L,y)=0$ and $K(x,0)=K(x,L)=0$. With these boundary conditions, the Helmholtz equation can be solved by separation of variables, obtaining $h = \alpha_{nm} \sin q_n (x-x_0) \sin q_m (y-y_0)$, where $q_n = n \pi/L_x$, $q_m = m \pi/L_y$, $x_0 = \mathrm{mod}(n,2) L/(2n)$ and $y_0 = \mathrm{mod}(m,2) L/(2m)$, $n,m \in \mathbb{N}$ and with $\bar{\Sigma} = q_n^2 +q_m^2$. The amplitude is given by $\alpha_{nm}^2 = (8/ \bar{\Sigma}) (\Delta A/A_p)$; it is proportional to the excess area $\Delta A = A - A_p$, and inversely proportional to $\Sigma$ (or the squared sum of the wave numbers), which depends on $\frm$. The first four modes are plotted in Fig. \ref{fig:3}.
At second-order the total bending energy can be written as $\bar{H}_B \approx \bar{\Sigma}^2/2 \int \rmd x \rmd y h^2$, whereas the total magnetic energy reads $\bar{H}_M \approx -\bar{\frm}/2 \int \rmd x \rmd y (\nabla h)^2$. Integrating, we get that the total energy is $\bar{H} \approx \bar{\Sigma}/2 (\Delta A - \bar{\frm} A_p/4)$, which is proportional to $q_n^2+ q_m^2$. Thus for fixed $\frm$, $A_p$ and $\Delta A$ the ground state corresponds to $n=m=1$.  If $\frm>0$ ($\frm<0$), $H$ increases (decreases). The total force exerted by the membrane along the edges at $x= \pm L_x/2$ is $ \bfF = \pm L_y \left[\left(-\bar{\sigma} + \bar{\frm}/2 \right) q_n^2/2 -\bar{\sigma} \right] \hat{\bfx}$; likewise, the total force along edges at $y= \pm L_y/2$ is $\bfF = \pm L_x \left[\left(-\bar{\sigma} + \bar{\frm}/2 \right) q_m^2/2 -\bar{\sigma} \right] \hat{\bfy}$. In both cases, if $\frm > 2 \sigma$ ($\frm < 2 \sigma$), the first term (arising from a combination of the tangential magnetic stress and the normal bending stress) is positive, so it represents a compressive (tensile) force. For $\sigma>0$ ($\sigma<0$) the second term represents a homogeneous tensile (compressive) force.

\subsection{Cylinders}

\begin{figure}[htbp]
\subfigure[]{\includegraphics[width=0.4\columnwidth]{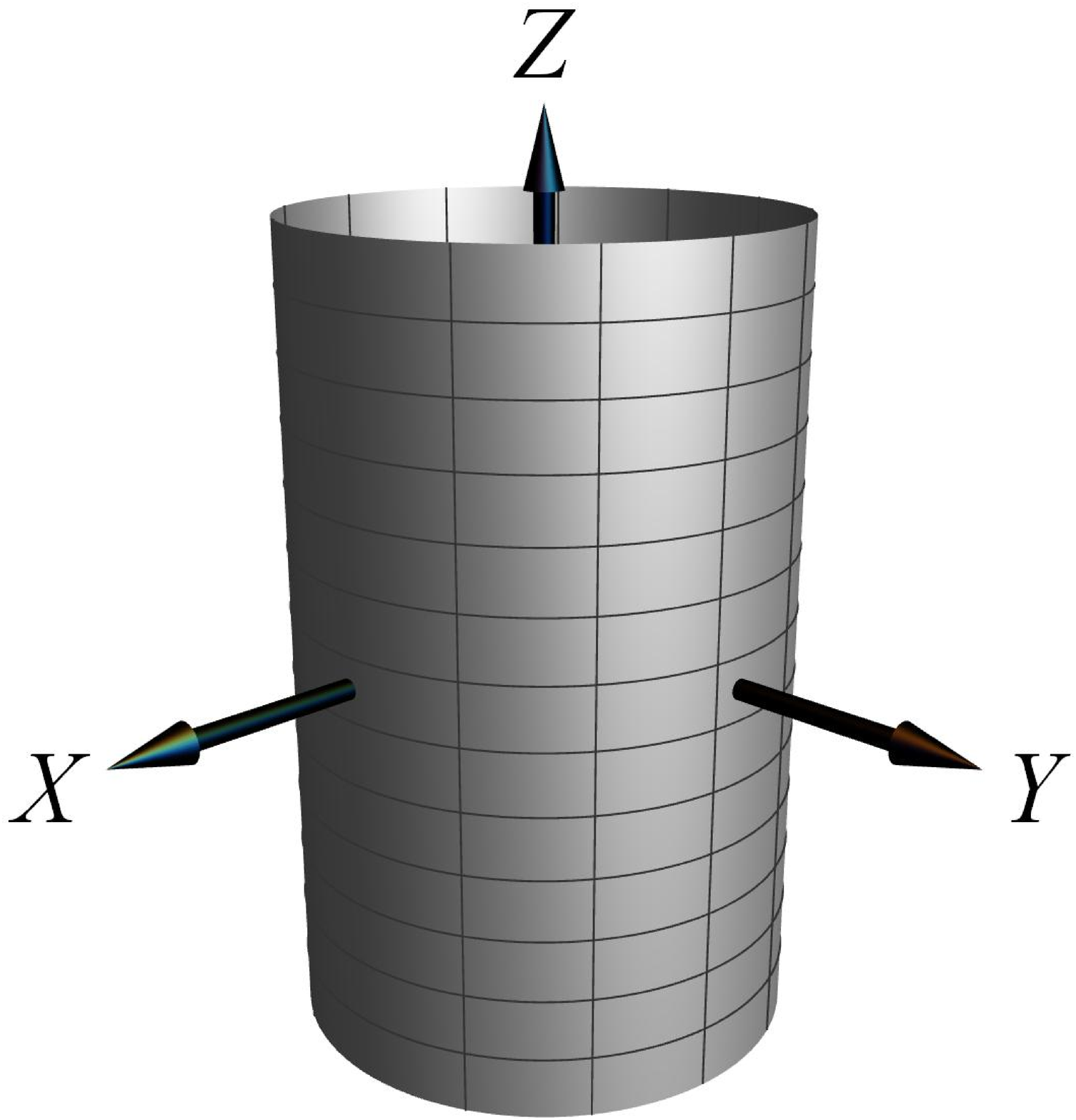}}
\subfigure[]{\includegraphics[width=0.4\columnwidth]{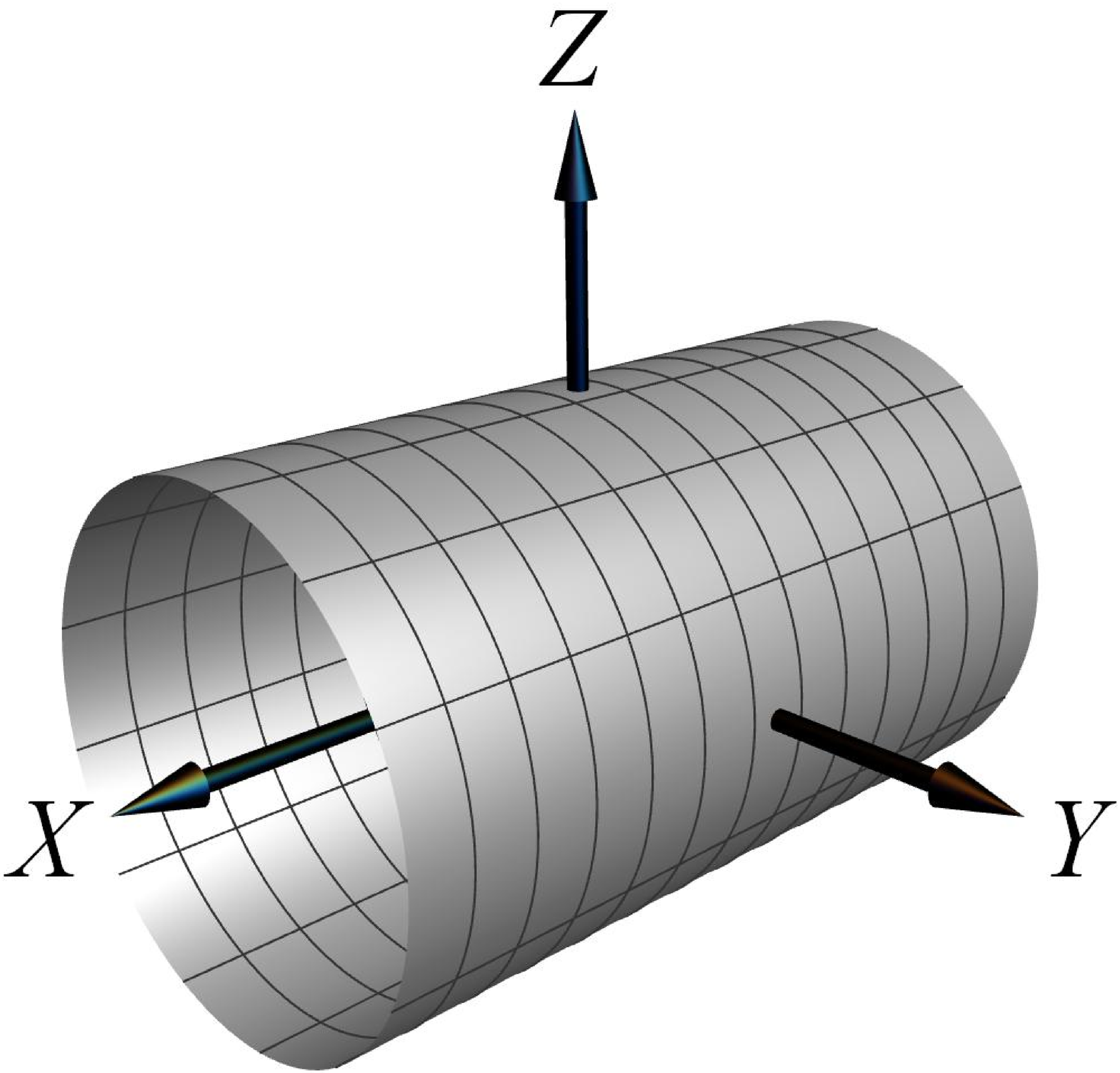}}
\caption{Cylinders oriented (a) along and (b) orthogonally to the axis of precession.}
\label{fig:4}
\end{figure}

\begin{figure}[htbp]
\includegraphics[width=0.5\columnwidth]{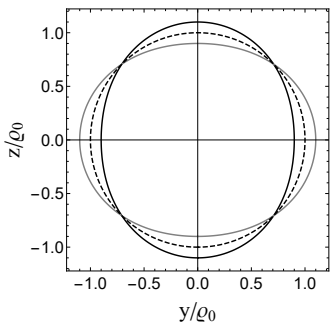}
\caption{Deformations of a circular cylinder ($\frm = 0$) oriented orthogonally to the precession axis $Z$ (dashed line). For $\frm>0$ its cross section elongates along the precession axis (black line), whereas for $\frm<0$ it gets squeezed in the orthogonal direction (gray line). The amplitude of the deformation has been exaggerated for illustration purposes.}
\label{fig:5}
\end{figure}

We now consider a cylinder of length $L$ aligned with the precession axis $Z$; see Fig. \ref{fig:4}(a).
The vectors tangent to the meridians are aligned with the precession axis, so their projection is constant, whereas the vector tangent to the parallel and the normal are orthogonal to it. Furthermore, $K$ is constant and $K_G$ vanishing. Thus, the EL Eq. (\ref{eq:ELmagmemb}) simplifies to $\bar{\mathcal{E}} = (-K^2/2 +\bar{\sigma}-\bar{\frm}/2) K = 0$, which is satisfied if the radius of the cylinder in equilibrium is $\varrho_e =1/K= 1/ \sqrt{2 \bar{\sigma} - \bar{\frm}}$,  ($\sigma > \frm/2$). Hence, for $\frm>0$ ($\frm < 0$), $\varrho_e$ is larger (smaller), than the corresponding radius of the purely elastic case. The total energy is $\bar{H}=\pi L(1/\varrho_e - \bar{\frm} \varrho_e)$, which is decreased (increased) for $\frm >0$ ($\frm <0$). The total force on a boundary parallel is $\bfF = -2 \pi/\varrho_e \, \hbfz$, so it is under tension (the membrane is pulling downwards), whereas the stress on a meridian vanish \cite{Deserno2015}. $\bfF$ is also modified in the presence of a magnetic field, for it depends on $\frm$ through $\varrho_e$.
\\
If we now consider a cylinder whose axis is orthogonal to the precession axis $Z$, say along the $X$ axis as shown in Fig. \ref{fig:4}(b), it is only solution in the purely elastic case with $\frm = 0$. This is because the magnetic contributions, proportional to the projections of the parallel's tangent and the normal onto the precession axis, vary along its circumference, while the bending contribution is constant, so the EL Eq. (\ref{eq:ELmagmemb}) cannot be satisfied everywhere. In contrast to the previous case of the vertical cylinder, where the equilibrium radius is determined for a given $\frm$, here we begin with a cylinder of fixed radius (and area) and we investigate how it gets deformed when $\frm \neq 0$. We assume that for a small magnetic modulus, $ \frm_1 \ll 1$, the membrane is still a generalized cylinder with $K_G=0$, so that we can examine linearized solutions about a circular cylinder of radius $\varrho_0$, which can be described in the cylindrical Monge representation as $\varrho(\vartheta) = \varrho_0 + \varrho_1 (\vartheta)$, where $\varrho_1 \ll 1$ is a small perturbation and $\vartheta$ is the azimuthal angle on the parallel measured from the precession axis $Z$. We also expand $\sigma = \sigma_0 + \sigma_1$, with $\sigma_1 \ll 1$. The first-order correction to the initial mean curvature $K_0=1/\varrho_0$ is $K_1 = -(\partial^2_\vartheta \varrho_1 + \varrho_1) /\varrho_0^2$. Inserting these results in the EL Eq. (\ref{eq:ELmagmemb}), we find that at lowest order it determines $\varrho_0 = 1/ \sqrt{2 \bar{\sigma}_{(0)}}$, whereas to first-order it reads
\begin{equation}
(\partial^2_{\vartheta} +1)^2 \varrho_1 = \varrho _0^3 \left(\frac{\frm_1}{4} (3 \cos 2 \vartheta + 1)-\sigma_1\right)\,. 
\end{equation}
The general solution of this differential equation is
\begin{eqnarray}
\varrho_1 &=& (a_1 \theta + b_1) \sin \vartheta + (c_1 \vartheta + d_1) \cos \vartheta \nonumber \\
&+&\frac{\varrho_0^3}{4} \left( \frac{\bar{\frm}_1}{3} \cos 2 \vartheta + \bar{\frm}_1 - 4 \bar{\sigma}_1 \right)\,,
\end{eqnarray}
with $a_1$, $b_1$, $c_1$, and $d_1$ constants of integration. Periodicity implies $a_1=c_1=0$ and we assume that solutions possess left-right symmetry, $\varrho_1(\vartheta) = \varrho_1(-\vartheta)$, as well as up-down symmetry, $\varrho_1(\vartheta) = \varrho_1(\pi-\vartheta)$, which entail $b_1=d_1=0$. Furthermore, the fixed area constraint requires the vanishing of the constant term (representing a dilation of the radius), so $\sigma_1 = \frm_1/4$. Finally the admissible deformation (preserving area to first-order) of the radius is $\varrho_1 = (\varrho_0^3  \bar{\frm}_1/12)  \cos 2 \vartheta$. Thus, for $\frm_1>0$ ($\frm_1<0$) the cross section of the cylinder elongates (flattens) along (orthogonally to) the precession axis; see Fig. \ref{fig:5}.
There is no first-order correction to the total energy of the cylinder, $H_1 = 0$, so at this order we have $H \approx \pi L/\varrho_0$. The force on a parallel is $\bfF = -2 \pi (1/\varrho_0 + \varrho_0 \frm_1/4) \hat{\mathbf{x}}$, thus, for $\frm_1 >0$ ($\frm_1 <0$) the first-order correction introduces tension (compression). There is no first-order correction to the total force on a meridian, so it vanishes at this order even in the deformed configurations.

\subsection{Helicoid}

\begin{figure}[htb]
\includegraphics[width=0.5\columnwidth]{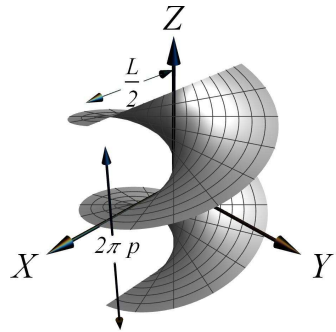}
\caption{A helicoid of pitch $2 \pi p$ bounded by two straight lines of length $L$ and two helices of radius $L/2$.} \label{fig:6}
\end{figure}

We know that helices minimize the sum of bending and magnetic linear energies \cite{Dempster2017, Vazquez2017}, and since the helicoid can be foliated by helices, we might expect it to be a critical point of $H$. The helicoid is a ruled minimal surface with glide rotational symmetry characterized by its pitch $2 \pi p$; see Fig \ref{fig:6}.
For a minimal surface, with $K=0$, the EL Eq. (\ref{eq:ELmagmemb}) reduces to $\mathcal{E}_{MS} = \frm K^{ab}\, \rme_a^z \, \rme_b^z = 0$. Let us consider a helicoid whose axis is aligned with the precession axis $Z$. Parametrizing it in cylindrical coordinates, by the distance along rulings and the azimuthal angle, we have that $K^{ab}$ is antidiagonal. Furthermore, terms in $\mathcal{E}_{MS}$ proportional to off-diagonal components of $K^{ab}$ vanish because the tangents along rulings are orthogonal to the precession axis, so the EL equation is satisfied. We consider that the helicoid is bounded at the top and bottom by two parallel straight lines of length $L$, and laterally by two helices completing one period. The total energy of such helicoid arises from the magnetic contribution and increases monotonically with $p$, $H=-2 \pi \frm p^2 \arcsinh \chi$, where $\chi = L/(2 |p|)$. The total forces on the boundary helices vanish, but the top and bottom boundary lines are subject to total forces along the helical axis,
\begin{equation}
\bfF = \mp |p| \left( 2 (\sigma - \frm ) \, \arcsinh \chi + \frac{\frm \, \chi}{\sqrt{\chi^2+1}} \right) \hbfz \,.
\end{equation}
For $\sigma > \frm > 0$ ($\sigma < \frm < 0$) the membrane is under tension (compression). We see that in this case the magnetic contribution not only rescales $\sigma$, but also introduces another dependence on $\chi$. 
 
\section{Discussion and conclusions}

We have extended the framework of elastic membranes to the study of flexible paramagnetic membranes. Unlike two-dimensional arrays of free paramagnetic particles, such membranes are able to resist shear forces and to buckle out of plane. To include the paramagnetic attribute of the membrane, we introduced an energy density representing the dipolar interactions between nearest neighbors. We have shown that such magnetic energy density is suitable to describe the main feature of paramagnetic arrays: alignment along the precession axis for small precession angles and flattening orthogonally to the precession axis for large precession angles. Minimization of the bending and magnetic energies permitted us to determine their equilibrium shapes and to quantify how they differ with their purely elastic counterparts. 
\\
Besides determining their shapes we have also presented the part of the surface stress tensor stemming from the magnetic dipolar interactions, which permitted us to analyze the underlying stresses shaping them. This information allows for a complete characterization of their conformations, for even when two membranes possess similar geometries, they well might be under very different forces. We have seen that even if paramagnetic membranes adopt shapes very close to those of purely elastic membranes, the magnetic field introduces additional stresses on them. This point is best illustrated on minimal surfaces, which regarded as elastic membranes are only under homogeneous tangential tension, whereas the paramagnetic membranes are subject to inhomogenous tangential and normal stresses.
\\
For the sake of simplicity we have considered only dipolar interactions of homogeneous arrays of beads, but several refinements to our model are possible. For instance one could take into account the long range nature of dipolar interactions and consider interactions beyond nearest neighbors, which may prevent the local magnetic field of the beads to precess at the magical angle, so that quadrupolar interactions never become dominant. One could also consider arrays of anisotropic paramagnetic beads, introducing anisotropies not only in the magnetic energy, but also in the bending energy. Such considerations would certainly complicate the analysis of their shapes and stresses, but they might also give rise to interesting phenomena not captured by our model.
\\
Examination of shapes in the nonlinear regime can be achieved by exploiting the residual symmetries of the total energy. For instance, axial symmetry about the precession axis provides a first integral of the EL equation, which will permit us to analyze how spheres or Clifford Tori (it can be shown that neither one is a critical point of the total energy) are modified by a magnetic field. As in the case of elastic membranes, further examination of general configurations will require numerical analysis and/or molecular dynamics simulations, which would reveal features with many more potential applications in actuation and soft robotics.

\section*{Acknowledgments}

We thank Joshua M. Dempster for interesting us in this problem. We have benefited from conversations with Chase Brisbois. This work was supported by the Center for Bio-Inspired Energy Science, which is an Energy Frontier Research Center funded by the U.S. Department of Energy, Office of Science, Office of Basic Energy Sciences under Award No. DE-SC0000989. P.V.M. acknowledges support by CONACYT Grant No. FORDECYT 265667.

\begin{appendix}

\section{Derivation of the magnetic energy density} \label{App:derHM}

We consider a membrane consisting of a homogeneous two-dimensional array of paramagnetic beads connected by elastic linkers in a magnetic field $\mathbf{H}$. In the high temperature and saturation magnetization regime, the induced dipole moment of each bead has magnitude $\mu$ and all are aligned with the magnetic field, so $\bm{\mu}_i = \bm{\mu} = \mu \, \hat{\bm \mu}$. Thus, the interaction energy of two paramagnetic beads at positions $\bfx_i$ and $\bfx_j$ is \cite{GriffithsBook, JacksonBook}
\begin{equation} \label{eq:dipEn}
U_{ij} = \frac{\mu_0 \mu^2}{4 \pi \, |\bfx_{ij}|^3} \left(1 - 3  
(\hat{\bm \mu} \cdot \hat{\bfx}_{ij})^2 \right)\,, 
\quad \bfx_{ij} = \bfx_j - \bfx_i\,.
\end{equation}
We consider as a unit cell the region spanned by four beads with edges of length $\Delta l$ and $\Delta w$ and denote the separation between the centers of the beads $\bfx_i -\bfx_{i-1}$ along the two directions by $\Delta \bfx$ and $\Delta \bfy$.
The energy per unit area associated to this unit cell due to dipolar interaction between nearest neighbors is
\begin{eqnarray} \label{eq:dipsurfEnDens}
u &=& \frac{\mu_0}{4 \pi} \Bigg[\frac{1}{\Delta w} \left(\frac{\mu}{\Delta l^2}\right)^2 \left(1 - 3  \left(\frac{\Delta \bfx}{\Delta l} \cdot \hat{\bm \mu}\right)^2  \right) \nonumber \\ 
&&+ \frac{1}{\Delta l} \left(\frac{\mu}{\Delta w^2}\right)^2 \left(1 - 3  \left(\frac{\Delta \bfy}{\Delta w} \cdot \hat{\bm \mu}\right)^2  \right) \Bigg]\,.
\end{eqnarray}
Assuming that $\Delta l \ll \mathrm{L}$ and $\Delta w \ll \mathrm{L}$, with $\mathrm{L}$ the length scale of the membrane, we can consider the separation of beads along the two different directions as the corresponding arc lengths, $\Delta l \rightarrow s$ and $\Delta w \rightarrow t$, so that the array can be approximated by the surface passing through the centers of the beads. In this approximation we have that $\frac{\Delta \bfx}{\Delta l} \rightarrow \hat{\bf V}: = \frac{\rmd \bfX}{\rmd s}$ and $\frac{\Delta \bfy}{\Delta w} \rightarrow \hat{\bf W} : = \frac{\rmd \bfX}{\rmd t}$. Additionally, we assume that the membrane is isotropic, so the separation between beads along different directions is approximately equal, $\Delta l \approx \Delta w$, and the coefficients tend to the same constant value (with units of energy per unit area). Thus, with these considerations we can define the surface energy density
\begin{equation} \label{eq:dipensimp}
\calH_M = \frac{3 \mu_0}{4 \pi \Delta l} \left(\frac{\mu}{\Delta l^2}\right)^2  \left(\frac{2}{3} - \left(\hat{\bf V} \cdot \hat{\bm \mu}\right)^2 - \left(\hat{\bf W} \cdot \hat{\bm \mu}\right)^2  \right)\,.
\end{equation}
Furthermore, if the magnetic field is precessing with an angular velocity $\omega$ about the $Z$ axis at an angle $\vartheta$, the direction of the induced dipole moment is $\hat{\bm \mu} = (\cos \omega t \, \sin \vartheta, \sin \omega t  \, \sin \vartheta, \cos \vartheta)$ and in the quasistatic regime (fast precession 
frequency) the tangent vectors $\hat{\bf V}$ and $\hat{\bf W}$ can be regarded as constant. In one period the average of the squared scalar product of $\hat{\bm \mu}$ and ${\bf V}$ is ($V^i$, $i=x,y,z$ stand for the cartesian components of the vector)
\begin{eqnarray} \label{eq:avmdotT}
\left\langle (\hat{\bm \mu} \cdot \hat {\bf V})^2 \right\rangle &=& \frac{1}{2} \sin^2 
\vartheta (V^x{}^2 + V^y{}^2) + \cos^2 \vartheta \, V^z{}^2 \nonumber\\
&=& \frac{1}{2} \left(\sin^2 \vartheta - (1-3 \cos^2 \vartheta)V^z{}^2\right)\,,
\end{eqnarray}
and similarly for vector $\hat{\bf W}$. Therefore, the time-averaged surface energy density is
\begin{eqnarray} \label{eq:avlindensen}
\left \langle \calH_M \right\rangle &=& \frac{3 \mu_0}{4 \pi \Delta l} \left(\frac{\mu}{\Delta 
l^2}\right)^2  \nonumber \\
&&\left(\frac{2}{3} - \sin^2 \vartheta +\frac{1}{2} \left(1-3 \cos^2 
\vartheta \right) \left( V^z{}^2 + W^z{}^2\right)  \right) \nonumber \\
&=&\frm \left[\frac{1}{3} - \frac{1}{2} \left( V^z{}^2 + W^z{}^2 \right) \right] \,,
\end{eqnarray}
where in the last step we defined the magnetic modulus
\begin{equation}
 \frm = \frac{\mu_0}{4 \pi \Delta l} \left(\frac{3 \mu}{\Delta l^2}\right)^2 \, \left(\cos^2 \vartheta - \frac{1}{3}\right) \,.
\end{equation}
The constant term of the magnetic dipolar energy, $\frm/3$, just renormalizes the intrinsic surface tension, so we neglect it. 
The unit tangent vectors can be spanned in the tangent basis of the surface as $\hat{\bf V} = V^a \, \bfe_a$ and $\hat{\bf W} = W^a \, \bfe_a$, therefore this energy density can be expressed in terms of the tangent basis as
\begin{equation} \label{eq:areaEnDensspan}
\calH_M =  - \frac{\frm}{2} \left(V^a \, V^b + W^a \, W^b \right) \rme_a^z \, \rme_b^z \,.
\end{equation}
Furthermore, we assume that the unit cell is approximately a square, such that $\{{\bf V}, {\bf W}\}$ constitute an orthonormal tangent basis, whose completeness permits us to span the metric as $g_{ab} = V_a \, V_b + W_a \, W_b$. Using this identity the surface energy density can be recast as the covariant expression given by Eq. (\ref{eq:HM}) of the main text.

\section{Estimation of the magnetoelastic parameter} \label{App:gammavals}

To calculate possible values of $\gamma$ we employ experimental values of the material parameters of paramagnetic filaments. Regarding the membrane as a thin plate of thickness $d$, the bending modulus is given by \cite{LandauBook, Deserno2015}
\begin{equation}
 \frk = \frac{Y d^3}{12 (1-\nu^2)} \,,
\end{equation}
where $Y$ is the Young modulus and $\nu$ the Poisson's ratio, whose typical values are of the order $Y \approx 10^3 - 10^5 \mathrm{Pa}$, \cite{Biswal2003, Goubault2003, Gerbal2015, Hu2018} and $\nu \approx 10^{-1}$. The diameter of the beads is approximately $d = \Delta l \approx 1 \upmu \mathrm{m}$, so the bending modulus is of the order of $\frk \approx 10^{-16} -10^{-14} \mathrm{J}$. Common values of the magnetic susceptibility and the magnetic fields are $\chi \approx 1$ and  $\mathrm{H}\approx 10^3 -10^4 \mathrm{A}/\mathrm{m}$ \cite{Biswal2003, Goubault2003, Gerbal2015, Hu2018}, so the magnitude of the induced dipole of a bead of radius $a=d/2$ is $\mu = 4/3 \pi a^3 \chi \mathrm{H} \approx  10^{-15} - 10^{-14} \mathrm{A} \mathrm{m}^2 $, thus $\frm \approx 10^{-6} - 10^{-4} \mathrm{N}/\mathrm{m}$. From this estimations we find that $\ell =\sqrt{\frk/|\frm|} \approx 10-100 \upmu \mathrm{m}$. Typical length scales of this kind of systems are $L \approx 10 - 10^3 \upmu \textrm{m}$ \cite{Biswal2003, Goubault2003, Gerbal2015, Hu2018}, so $A \sim L^2 \approx 10^{-10} -10^{-6} \mathrm{m}^2$. Therefore, possible values of the magnetoelastic parameter are $|\gamma| = A /\ell^2 \approx 10^{-2}- 10^{4}$. 

\section{Stress tensor projections} \label{App:stressprojs}

Consider a region $D$ of the membrane bounded by a curve $\partial D$, with tangent $\bft=\rmt^a \bfe_a$ and outward conormal $\bfl= \bft \times \bfn = \rml^a \bfe_a$. The projection of the stress tensor onto $\bfl$, $\bff_\perp = \rml_a \bff^a$ represents the force per unit length exerted by $D$ on the neighboring region. This projection can be expressed in the Darboux frame $\{\bft,\bfl,\bfn\}$ as $\bff_{\perp} = f_{\perp \parallel }\bft + f_{\perp \perp} \bfl +f_\perp \bfn$, where the components are defined by
\begin{subequations} \label{def:projsfprp}
\begin{eqnarray}
\bar{f}_{\perp \parallel} &:=& \rml_a \rmt_b  f^{ab} = - K \tau_g \,,\\ 
\bar{f}_{\perp \perp} &:=& \rml_a \rml_b  f^{ab} = \frac{1}{2} \left( \kappa_{n \perp}^2 -\kappa_n^2 \right) -\bar{\sigma} \nonumber \\
&\phantom{:=}& \phantom{\rml_a \rml_b  f^{ab}} + \frac{\bar{\frm}}{2} \left( \rmt^z{}^2 + \rml^z{}^2 \right)   \,,  \\
\bar{f}_{\perp} &:=& \rml_a f^a = - \nabla_\perp K + \bar{\frm} \, \rml^z \rmn^z \,,
\end{eqnarray}
\end{subequations}
with $\kappa_n = \rmt^a \rmt^b K_{ab}$ and $\kappa_{n \perp} = \rml^a \rml^b K_{ab}$ the normal curvatures along and across the curve ($K = \kappa_n + \kappa_{n\perp}$); $\tau_g = - \rmt^a \rml^b K_{ab}$ is the geodesic torsion of the curve; $\nabla_\perp = l^a \nabla_a$.
\\
We see that the tangential bending stresses do not have a definite sign, so they can represent either compression ($+$) or tension ($-$). By contrast, the sign of tangential magnetic stress is given entirely by the sign of the magnetic modulus: for $\frm>0$ the magnetic field introduces a compressive stress on the membrane, and for $\frm<0$ a tensile stress. Due to the minus sign in front, $\sigma>0$ introduces tension and $\sigma<0$ compression.

\section{Surface Torque tensor} \label{App:torquetensor}

The surface torque tensor is given by \cite{CapoGuv2002, Guven2018}
\begin{equation}
 \bar{\bfm}^a = \bfX \times \bar{\bff}^a + \left[ K g^{ab} - \bar{\frk}_G \left(K^{ab} - K g^{ab}\right) \right] \bfe_b \times \bfn \,,
\end{equation}
Taking its covariant derivative we get 
\begin{equation} \label{divm}
\nabla_a \bfm^a = \bfX \times \nabla_a \bff^a + \frm \,  \rmn^z \, \hbfz \times \bfn\,,
\end{equation}
where we have used the Codazzi-Mainardi integrability condition $\nabla_a \left(K^{ab}-K g^{ab}\right)=0$, as well as the identity $\hbfz = g^{ab} \rme_a^z \bfe_b + \rmn^z \bfn$. While the first term vanishes in equilibrium on account of the conservation law of the stress tensor, the second term, representing a torque per unit area due to the magnetic field, does not vanish in general. Thus, the torque tensor is not conserved. However, its projection onto the precession axis, $\mathrm{m}^a = \bfm^a \cdot \hbfz$, is conserved in equilibrium, for we have $\nabla_a \mathrm{m}^a = \hbfz \cdot \bfX \times \nabla_a \bff^a $. This is a consequence of the rotational symmetry of $H$ about the precession axis. The torque per unit length exerted by a region $D$, is given by the projection 
\begin{equation}
 \bar{\bfm}_\perp = \rml_a \bar{\bfm}^a = \bfX \times \bar{\bff}_\perp - K \bft - \bar{\frk}_G \left(\kappa_n \bft -\tau_g \bfl \right) \,.
\end{equation}
Integrating the torque per unit area by a region $D$, given by Eq. (\ref{divm}), we find that the membrane is subject to a total torque about the precession axis, $\bfM = \int \rmd s \, \bfm_\perp =  \frm \,  \int \rmd A \, \rmn^z \, \hbfz \times \bfn$.

\section{Alternative expressions} \label{App:altexp}

Up to a constant term, the energy density Eq. (\ref{eq:HM}) could also be written as $\calH_M = g^{ab} \left( \bfe_a \times \hbfz \right) \cdot \left( \bfe_b \times \hbfz \right)$, or in terms of the antisymmetric tensor $\varepsilon_{ab} = \bfn \cdot (\bfe_a \times \bfe_b)$, as $\calH_M = g_{ab} \varepsilon^{ac} \varepsilon^{bd} \rme_{c}^z \rme_{d}^z$.
\\
Moreover, squaring the identity $\hbfz = g^{ab} \rme_a^z \bfe_b + \rmn^z \bfn$, we have $1 = g^{ab} \rme_a^z \rme_b^z + \rmn^z{}^2$. This relation permits us to express the magnetic energy density Eq. (\ref{eq:HM}) in a coordinate-free manner in terms of the squared projection of the unit normal vector onto the precession axis:
\begin{equation} \label{eq:HMalt}
\calH_M = \frac{\frm}{2} \, \rmn^z{}^2 \,.
\end{equation}
In this manner it becomes clear that $\calH_M$ does not depend on how the beads are distributed in the membrane. The constant term involving the magnetic modulus can be absorbed in the intrinsic surface tension, $\sigma \rightarrow \sigma - \frm/2$.
\\
Variation of the energy density (\ref{eq:HMalt}) leads to the following expressions of the magnetic parts of the tangential components of the stress tensor and of the EL equation
\begin{subequations}
 \begin{eqnarray}
 f^{ab}_M &=& - \frac{\frm}{2} \, \rmn^z{}^2 \, g^{ab} \,, \\
 \varepsilon_M &=& \frm \left( K ^{ab} \rme_a^z \rme_b^z -\frac{K}{2} \, \rmn^z{}^2 \right) \,,
\end{eqnarray}
\end{subequations}
which are equivalent to the corresponding expressions in Eqs. (\ref{eq:fab}) and (\ref{eq:ELmagmemb}), up to the redefinition of $\sigma$ mentioned above.

\section{Geometric quantities} \label{App:geomquant}

Here we present the expressions of the geometric quantities required to calculate the EL Eq. (\ref{eq:ELmagmemb}) for each geometry, and defined by ($a=1,2$),
\begin{subequations}
 \begin{align}
\mathbf{X}&=(X^1,X^2,X^3)\,, & \mbox{embedding functions} \\
\bfe_a &= \partial_a \bfX & \mbox{adapted tangent vectors} \nonumber \\
\bfn &= \frac{\bfe_1 \times \bfe_2}{|\bfe_1 \times \bfe_2|} & \mbox{unit normal vector} \\
g_{ab} & = \bfe_a \cdot \bfe_b & \mbox{metric tensor} \\
K_{ab} &= \bfe_a \cdot \partial_b \bfn & \mbox{curvature tensor}
 \end{align}
\end{subequations}
\\
\textit{Almost planar membranes}: ($\nabla h= \partial_x h \hat{\bfx} + \partial_y h\hat{\bfy}$)
\begin{subequations}
 \begin{eqnarray}
\bfX (x,y) &=& x \hat{\mathbf{x}} + y \hat{\mathbf{y}} + h(x,y) \hbfz\,; \\
\bfe_x &=& \hat{\mathbf{x}} + \partial_x h \hbfz \,, \quad \bfe_y = \hat{\mathbf{y}} + \partial_y h \hbfz \,, \nonumber \\
\bfn &=& \frac{1}{\sqrt{1+(\nabla h)^2}} \left( -\partial_x h \hat{\mathbf{x}}  -\partial_y h \hat{\mathbf{y}} + \hbfz \right)\,; \quad \\
g_{ab} &=& \delta_{ab} + \partial_a h \partial_b h  \,; \\
K_{ab} &=& \frac{-\partial_a \partial_b h}{\sqrt{1+(\nabla h)^2}} \,.
\end{eqnarray}
\end{subequations}
\textit{Vertical cylinder}: Cylindrical basis $\{ \hat{\bm{\varrho}}, \hat{\bm{\varphi}}, \hbfz\}$, with $\hat{\bm{\varrho}} = \cos \varphi \hat{\mathbf{x}} + \sin \varphi \hat{\mathbf{y}}$ and $\hat{\bm{\varphi}} = -\sin \varphi \hat{\mathbf{x}} + \cos \varphi \hat{\mathbf{y}}$;
\begin{subequations}
\begin{eqnarray}
\bfX (\varphi,z) &=& \varrho_0 \hat{\bm{\varrho}} + z \hbfz\,; \\
\bfe_\varphi &=& \varrho_0 \hat{\bm{\varphi}}\,, \quad \bfe_z = \hbfz \,, \quad \bfn = \hat{\bm{\varrho}}\, ; \\
g_{\varphi \varphi} &=& \varrho_0^2\,, \quad g_{\varphi z} = 0\,, \quad g_{zz}=1  \,; \\
K_{\varphi \varphi} &=& \varrho_0 \,, \quad  K_{\varphi z} = K_{zz}= 0\,.
\end{eqnarray}
\end{subequations}
\textit{Horizontal cylinder}: Cylindrical basis $\{ \hat{\mathbf{x}}, \hat{\bm{\vartheta}}, \hat{\bm{\varrho}} \}$, with $\hat{\bm{\vartheta}} = \cos \vartheta \hat{\mathbf{y}} -\sin \vartheta \hat{\mathbf{z}}$ and $\hat{\bm{\varrho}}(\vartheta) = \sin \vartheta \hat{\mathbf{y}} + \cos \vartheta \hat{\mathbf{z}}$;
\begin{subequations}
 \begin{eqnarray}
\bfX(x,\vartheta) &=& x \hat{\bfx} + \varrho(\vartheta) \hat{\bm{\varrho}}(\vartheta)\,; \\
\bfe_x &=& \hat{\mathbf{x}}\,, \qquad \bfe_\vartheta = \partial_\vartheta \varrho \hat{\bm{\varrho}} + \varrho \hat{\bm{\vartheta}}\,, \nonumber\\
\bfn &=& \frac{1}{\sqrt{\varrho^2 + (\partial_\vartheta \varrho)^2}} \left(\varrho \hat{\bm{\varrho}} - \partial_\vartheta \varrho \hat{\bm{\vartheta}}\right)\,; \\
g_{xx}&=&1\,, \quad g_{x\vartheta} = 0\,, \quad g_{\vartheta \vartheta} = \varrho^2 + (\partial_\vartheta \varrho)^2\,; \quad \\
K_{xx}&=& K_{x\vartheta} = 0\,, \nonumber \\
K_{\vartheta \vartheta} &=& \frac{\varrho  \left( - \partial^2_{\vartheta} \varrho + \varrho  \right) + 2 (\partial_\vartheta \varrho)^2}{\sqrt{\varrho^2 + (\partial_\vartheta \varrho)^2}} \,.
\end{eqnarray}
\end{subequations}
\textit{Helicoid}: Cylindrical basis $\{ \hat{\bm{\varrho}}, \hat{\bm{\varphi}}, \hbfz\}$, with $\hat{\bm{\varrho}} = \cos \varphi \hat{\mathbf{x}} + \sin \varphi \hat{\mathbf{y}}$ and $\hat{\bm{\varphi}} = -\sin \varphi \hat{\mathbf{x}} + \cos \varphi \hat{\mathbf{y}}$;
\begin{subequations}
 \begin{eqnarray}
\bfX(\varrho, \varphi) &=& \varrho \hat{\bm \varrho} + p \varphi \hbfz\,; \\
\bfe_\varrho & = & \hat{\bm \varrho}\,, \qquad \bfe_\varphi = \varrho \hat{\bm \varphi} + p \hbfz \,, \nonumber\\
\bfn &=& \frac{1}{\sqrt{\varrho^2 + p^2}} \left(- p \hat{\bm \varphi} + \varrho \hbfz \right)\,; \\
g_{\varrho \varrho} &=& 1 \,, \quad g_{\varrho \varphi} = 0\,, \quad g_{\varphi \varphi} = \varrho^2 + p^2\,; \\
K_{\varrho \varrho}&=& K_{\varphi \varphi} = 0\,, \quad K_{\varrho \varphi} = \frac{p}{\sqrt{\varrho^2 + p^2}} \,.
\end{eqnarray}
\end{subequations}

\end{appendix}

\bibliography{BibMagMemb}

\end{document}